\documentclass[10pt,journal,twocolumn]{IEEEtran}
\usepackage{amssymb,cite}
\usepackage{amsmath,multirow,threeparttable,cite,graphicx,times,subfigure,url,verbatim,color,multirow}
\usepackage[normalem]{ulem}

\definecolor{red}{rgb}{0.8,0,0}

\begin{document}

\title{In-Edge AI: Intelligentizing Mobile Edge Computing, Caching and Communication by Federated Learning}

\author{
    Xiaofei Wang$^{1}$,
    Yiwen Han$^{1}$,
    Chenyang Wang$^{1}$,
    Qiyang Zhao$^{2}$,
    Xu Chen$^{3}$,
    Min Chen$^{4*}$
    \\
    \scriptsize
    ${^1}$Tianjin Key Laboratory of Advanced Networking (TANK), School of Computer Science and Technology,
    Tianjin University, Tianjin, China.\\
    ${^2}$Huawei Technology, Shenzhen, P. R. China.\\
    ${^3}$School of Data and Computer Science, Sun Yat-sen University, Guangdong, Guangzhou, P. R. China.\\
    ${^4}$School of Computer Science and Technology, Huazhong University of Science and Technology, Wuhan, Hubei, P. R. China.\\
    *Corresponding author.
    \thanks{@ 2019 IEEE.  Personal use of this material is permitted.  Permission from IEEE must be obtained for all other uses, in any current or future media, including reprinting/republishing this material for advertising or promotional purposes, creating new collective works, for resale or redistribution to servers or lists, or reuse of any copyrighted component of this work in other works.}
}

\markboth{IEEE NETWORK MAGAZINE, VOL. XX, NO. YY, MONTH XXXX}{}
\maketitle

\begin{abstract}

Recently, along with the rapid development of mobile communication technology,
edge computing theory and techniques have been attracting more and more
attentions from global researchers and engineers,
which can significantly bridge the capacity of cloud and requirement of devices by the network edges,
and thus can accelerate the content deliveries and improve the quality of mobile services.
In order to bring more intelligence to the edge systems, compared to traditional optimization methodology,
and driven by the current deep learning techniques,
we propose to integrate the Deep Reinforcement Learning techniques
and Federated Learning framework with the mobile edge systems,
for optimizing the mobile edge computing, caching and communication.
And thus, we design the ``In-Edge AI'' framework in order to intelligently utilize
the collaboration among devices and edge nodes
to exchange the learning parameters for a better training and inference of the models,
and thus to carry out dynamic system-level optimization and application-level enhancement
while reducing the unnecessary system communication load.
``In-Edge AI'' is evaluated and proved to have near-optimal performance
but relatively low overhead of learning,
while the system is cognitive and adaptive to the mobile communication systems.
Finally, we discuss several related challenges and opportunities
for unveiling a promising upcoming future of ``In-Edge AI''.

\end{abstract}


\section{Introduction}
\label{sec:intro}
With the increasing quantity and quality of rich multimedia services over mobile networks, there has been a huge increase in the traffic and computation for mobile users and devices
over recent years, which imposes a huge amount of workload
on today's already-congested backbone networks and the mobile networks.

Naturally, the emerging idea of Mobile Edge Computing (MEC) is proposed \cite{edge_etsi_report}
as a novel paradigm for easing the burden of backbone networks by pushing the computation/storage resources
to the proximity of the User Equipments (UEs). On the other hand,
MEC circumvents the long propagation delays introduced by transmitting
data from mobile devices to remote cloud computing infrastructures,
and hence is able to support latency-critical mobile and Internet of Things (IoT) applications.
Specifically, edge nodes, i.e., base stations equipped with computation/storage capability,
could deal with the computation and content requests of UEs,
and consequently this scheme improves the Quality-of-Service (QoS) of Mobile Network Operators (MNOs) and the Quality-of-Experience (QoE) of UEs and relieves the load of backbone networks,
and pressure of clouds (data centers) \cite{cacheintheair}.

Fulfilling the requirement of QoE of UEs is not a trivial even by virtue of MEC.
The key and difficult point lies in that the computation offloading requires wireless data transmission
and might bring about the congestion of wireless channels,
and hence raises the decision making or optimization problem on the whole communication
and computation integrated system, i.e., how to jointly allocate communication resources
and computation resources of edge nodes.

Several pioneer works have been proposed and realize quite good results in their assuming settings
based on convex optimization, game theory and so on  \cite{CHENMIN_MEC} \cite{COMST_MEC_COMMUNICATION} .
Nevertheless, considering the particular use cases in MEC,
these optimization methods may suffer from the following issues:
\textbf{\emph{1) Uncertain Inputs:}} they assume the that some key information factors are given as inputs,
but actually some of them are difficult to obtain due to variant wireless channels and privacy policies;
\textbf{\emph{2) Dynamic Conditions:}} dynamics of the integrated communication
and computation system are not well addressed;
\textbf{\emph{3) Temporal Isolation:}} most of them do not consider the long-term effect of
current decision on resource allocation except for Lyapunov optimization,
viz., in a highly time-varying MEC system, most of proposed optimization algorithms is
optimal or close-to-optimal only for a snapshot of the system.
In a word, the problem existed in the resource allocation optimization of the MEC system
is ``\textbf{lack of intelligence}''.

In view of the increasing complexity of mobile networks, e.g., a typical 5G node is expected to have 2000 or more configurable parameters, a recent new trend is to optimize wireless communication by Artificial Intelligence (AI) techniques
\cite{COMST_MEC_COMMUNICATION} \cite{COMST_DL_IWN},
include but not limited to the application of AI for Physical Layer (PHY),
Data Link Layer, and traffic control \cite{dl_routing}. Particularly, related studies on edge computing and caching,
such as \cite{JSTSP_RL_CACHING} \cite{TVT_INTEGRATED_NETWORKING_DRL},
have shown that reinforcement learning \cite{BOOK_RL_INTRODUCTION}
(include Deep Q-Learning \cite{NATURE_2015_DQN} in Deep Reinforcement Learning)
has the potential to be effective in joint resource management.
But \cite{JSTSP_RL_CACHING} is based on Q-Learning, which is not feasible for the practical MEC system
where the state-action space is tremendous and the focus of \cite{TVT_INTEGRATED_NETWORKING_DRL} is on vehicular networks.
Besides, none of these works has thought over
1) in what form the training data shall be gathered
(whether in a distributed or a centralized way),
2) where the reinforcement learning agent should be placed and trained
(whether in UEs, edge nodes or remote cloud infrastructures),
3) how the update process of reinforcement learning agents should be proceeded and collaborated,
and
4) the privacy protection of training data.

\begin{figure*}[t]
\centering
\includegraphics[width=18cm]{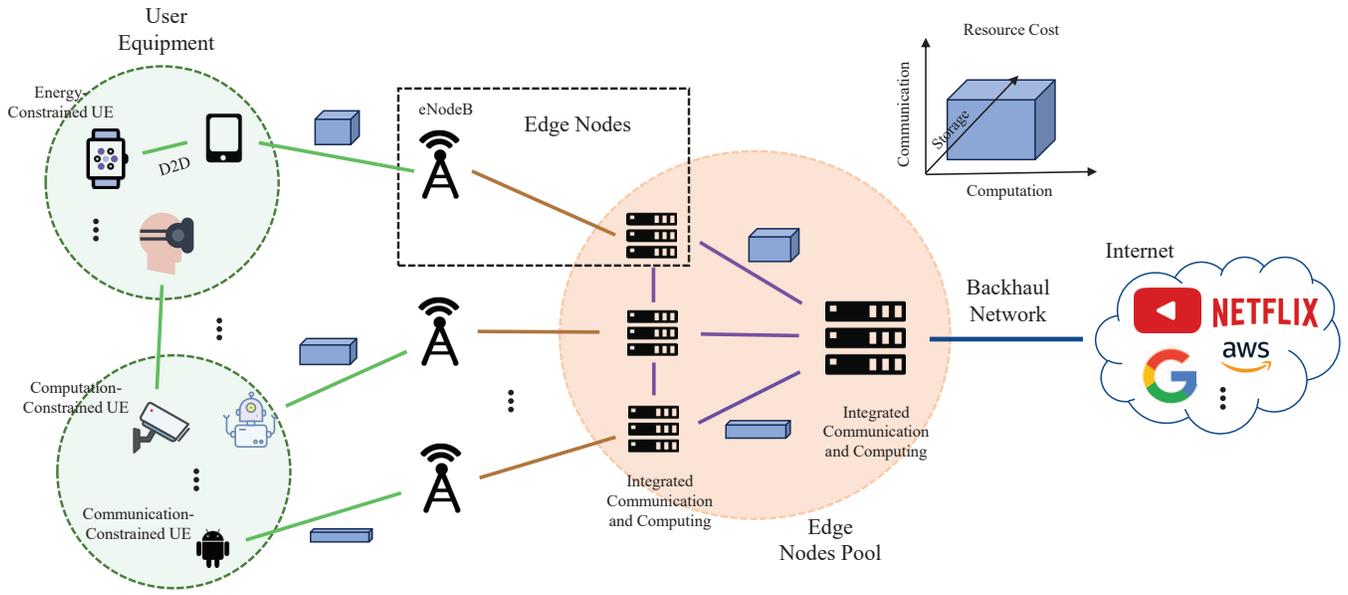}
\caption{Framework of AI-supported mobile edge system with cognitive ability}
\label{fig_1_illustrationRDL}
\end{figure*}

Therefore, in this article, we use Deep Reinforcement Learning (DRL) to jointly manage the communication and computation resources. And both cases of computation offloading and edge caching among the MEC system (illustrated in Fig. \ref{fig_1_illustrationRDL}) is discussed.
In addition, Federated Learning \cite{GOOGLE_FL} is introduced as a framework for
training DRL agents in a distributed manner while
1) largely reducing the amount of data which should be uploaded via the wireless uplink channel,
2) reacting cognitively to the mobile communication environment and conditions of cellular networks,
3) adapting well with heterogeneous UEs in a practical cellular network,
and 4) preserving the personal data privacy,

To the best of our knowledge, we are the first group to
study the application of DRL coupled
with Federated Learning for intelligent joint resource management
of communication and computation in MEC systems.
Our contribution can be summarized as follows:
\begin{itemize}
  \item 1) we discuss the methodology of utilizing DRL (specifically, Deep Q-Learning), and Distributed DRL for optimizing the Edge Caching and Computation;
  \item 2) we propose ``In-Edge AI'' framework to further
  utilize the ``Federated Learning'' for a better deployment
  of intelligent resource management in the MEC system;
  \item 3) we provide proof-of-concept evaluation and verify that
  the proposed scheme has advantages on the balance of performance and cost.
\end{itemize}
In addition, we also discuss opportunities and challenges
to hopefully unveil the upcoming future of edge architecture supporting various AI-based applications.

\section{Optimizing the Edge by DRL}
\label{sec:AIForEdge}
\begin{figure*}[t]
\centering
\includegraphics[width=15 cm]{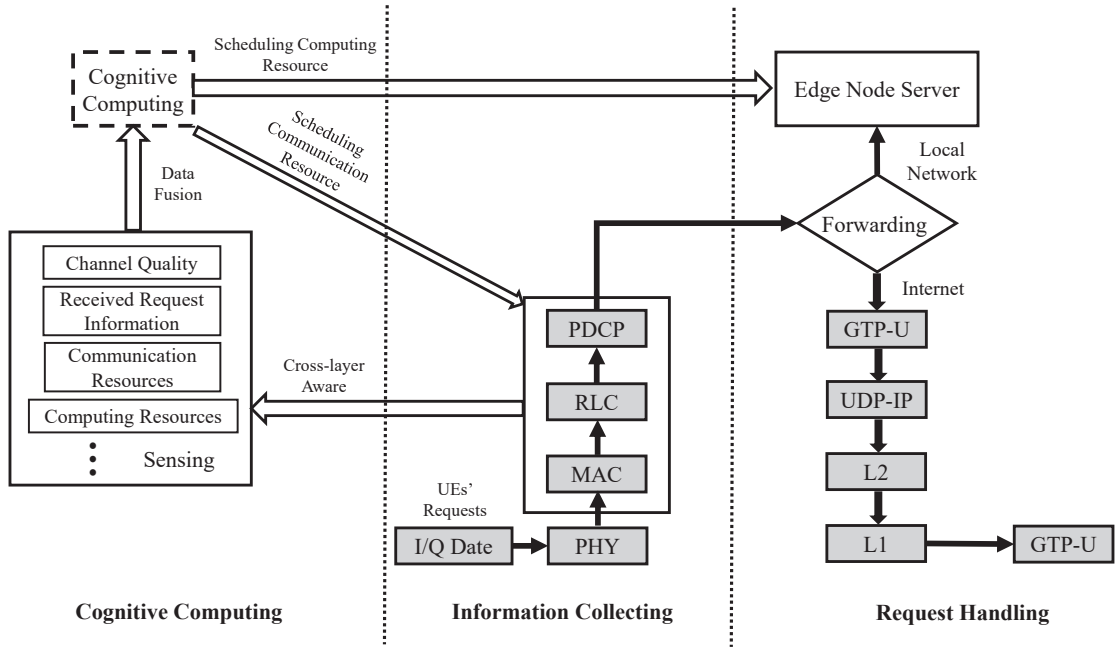}
\caption{Procedure of utilizing cognitive computing in mobile edge system among protocol stacks}
\label{fig_2_aicanoptimizeedge}
\end{figure*}

We intend to use AI techniques (particularly DRL) as the method of cognitive computing for building an intelligentizing mobile edge computing, caching and communication system. The cognitive process among protocol stacks of wireless communication is given as Fig. \ref{fig_2_aicanoptimizeedge}, where we partition the whole process into three main parts.
\begin{itemize}
\item \textbf{Information Collecting:} Sense and collect the indispensable observing data for cognitive computing among the MEC system, including but not limit on the usage of communication and computation resources, wireless environments and intensities of UEs' requests;
\item \textbf{Cognitive Computing:} By making use of the observing data of the system, cognitive computing is performed to fuse the massive observed data and further give the decision of scheduling;
\item \textbf{Request Handling:} The MEC system deals with the request of UEs on the basis of the scheduling decision given by cognitive computing.
\end{itemize}

In this section, two representative use cases in the MEC system are investigated.

\subsection{DRL over the MEC System for Caching}
\label{sec:AIForEdgeCaching}
Recently, we have observed the emergence of promising mobile content caching and delivery techniques, by which popular contents are cached in the intermediate servers (or middleboxes, gateways or routers) so that demands from users for the same content can be accommodated easily without duplicate transmissions from remote cloud servers, and hence significantly reducing redundant traffic.

Thereinafter, we focus on the scenario of caching contents in edge nodes. In the MEC system depicted by Fig. \ref{fig_1_illustrationRDL}, there is a library of $F$ popular content files, denoted as $\mathcal{F}=\{1,...,F\}$, that all mobile users may request in the system. The content popularity is defined as $(P_\mathrm{f})_{F\times 1}$, which is the probability distribution of content requests from all users. Content popularity indicates the common interests of all users in the network. In related works, the content popularity is always described by MZipf distribution. Moreover, for simply asserting the efficiency of DRL in edge caching, we assume that the content popularity changes slowly and all contents have the same size. For each request, the DRL agent in the edge node can make a decision to cache or not cache, and if yes, the agent determines which local content shall be replaced. We assume that all content popularity, user preference and average arrival rate of requests are static during a relatively long period. We model the cache replacement problem in all edge nodes as a Markov Decision Process (MDP) and use DRL to solve it, and the results will be shown in Section \ref{sec:evaluation}.

\subsection{DRL over the MEC System for Computation Offloading}
\label{sec:AIForEdgeComputing}
\subsubsection{Communication Model}
As illustrated in Fig. \ref{fig_1_illustrationRDL}, considering an environment where a set of UEs $\mathcal{N}=\{1,...,N\}$ is covered by a set of base stations $\mathcal{B}=\{1,...,B\}$, UEs could choose to offload their intensive computation  tasks to an edge node via the wireless channel or execute these tasks locally. There are $M$ wireless channels and the set $\mathcal{M}=\{1,...,M\}$ denotes channels of one base station. Specifically, among the decision choice $a_n\in \{0\} \bigcup \mathcal{M}$, UE $n$ could choose to offload the computation to the edge via a wireless channel $a_n$ or compute its tasks locally as a decision $a_n=0$. For the purpose of simulating the variation of wireless channels, the channel gain state between a UE and a base station (belongs to an edge node) is independently picked from a finite state space, by which the channel state transitions are modelled as a finite-state discrete-time Markov chain. In this wireless scenario, the achievable data rate could be evaluated by Shannon-Hartley theorem.

\subsubsection{Computation Model}
In order to portrait the long-term effects within the MEC system, computation tasks are generated according to Bernoulli distribution across the time horizon.  A computation task is represented by $(\mu, \nu)$, where $\mu$, $\nu$ denote the size of computation input data (in bits) and the total number of CPU cycles needed to complete the computation task, respectively. All these generated tasks are stored in a task queue and executed sequentially on the UE or edge node according to a FIFO (First-in First-out) principle. When the task is executed locally, the computation execution time of it is given as $d_L=\nu / f_L$, where $f_L$ is the the computation capability (i.e., CPU cycles per second) of the UE and determined by the amount of energy the UE decides to allocate. And when the task is scheduled to be executed on the edge node, the execution time of this offloaded task can be calculated as $d_E=\nu / f_E$, where $f_E$ is the computation capability the edge node allocated to the UE and the formula $f_E \gg f_L$ holds (the computing performance of edge nodes is much stronger than the UE's).

\subsubsection{Problem Formulation for Computation Offloading}
For efficiently performing computation offloading over the MEC system, the UE shall make a joint communication and computation resource allocation decision in terms of a control action $(c, e)$, where $c \in \{0\} \bigcup \mathcal{M}$ is the computation offloading decision denoting the UE chooses to execute the task locally ($c=0$) or to offload the task via which wireless, and $e$ denotes the amount of allocated energy for wireless communication and locally computation. In the MEC system, our attention in this section is focus on how to improve the task executing experience (viz., QoE) of UEs.

High complexities of the MEC system will be introduced especially when dealing with task offloading problems and the difficulty of acquiring the global necessary information when involving massive UEs. Hence, DRL algorithm such as Deep Q-Leaning is utilized here as the agent, which handles the joint control action of the UE.
The whole problem could be summarized as that the UE decides a joint wireless channel selection and energy allocation decision according to a stationary control policy $\mathbf{\Phi}=(\Phi_\mathrm{c}(\mathbf{\Upsilon}), \Phi_\mathrm{e}(\mathbf{\Upsilon}))$, while it keeps observing the network state $\mathbf{\Upsilon}$ which involves the task queuing state, the cumulative energy consumption of the UE, the occupying wireless channel of the UE and qualities of all wireless channels. In addition, we define an immediate utility function $u(\mathbf{\Upsilon}, (c, e))$ to evaluate the QoE of UEs, which is inversely proportional to the execution delay of tasks (include the wireless transmission delay and the computation delay), the task queuing delay, the energy consumption of the UE and the count of task dropping and failing. By taking advantage of Deep Q-Learning and its improved version, e.g., Double DQN \cite{google_double_dqn}, the control policy $\mathbf{\Phi}=(\Phi_\mathrm{c}(\mathbf{\Upsilon}), \Phi_\mathrm{e}(\mathbf{\Upsilon}))$ could be trained and achieves increasing the utility of UEs for the long-term performance optimization.

\section{In-Edge AI with Federated Learning}
\label{sec:AIInEdge}
In Section \ref{sec:AIForEdge}, two use cases of DRL in the MEC system are put forward. However, one key challenge is still pending in practical, namely the deployment of DRL agent. Taking the computation offloading use case in Section \ref{sec:AIForEdgeComputing} as example, \emph{if the DRL agent is trained on edge nodes or remote cloud servers just as depicted in Fig. \ref{fig_4_tree}(a)}, due to the wireless communication characters of the MEC system, three deficiencies existed: 1) the training data is large in quantity when considering massive UEs, and it will increase the burden of uplink wireless channels; 2) the training data which should be uploaded to edge nodes or cloud is privacy-sensitive, and it might cause potential privacy accidents; 3) if training data is transformed for privacy consideration, server-side proxy data is less relevant than on-device data. And \emph{if the DRL agent is trained on the UE individually}, there are still another two defects: 1) the computation capability of a UE is relatively weak, and it will consume long time or is even impossible to train the DRL agent on large-scale data; 2) the training process of a DRL agent may introduce extra energy consumption of a UE.

\begin{figure*}[!hhhhhhhhhht]
\centering
\includegraphics[width=14cm]{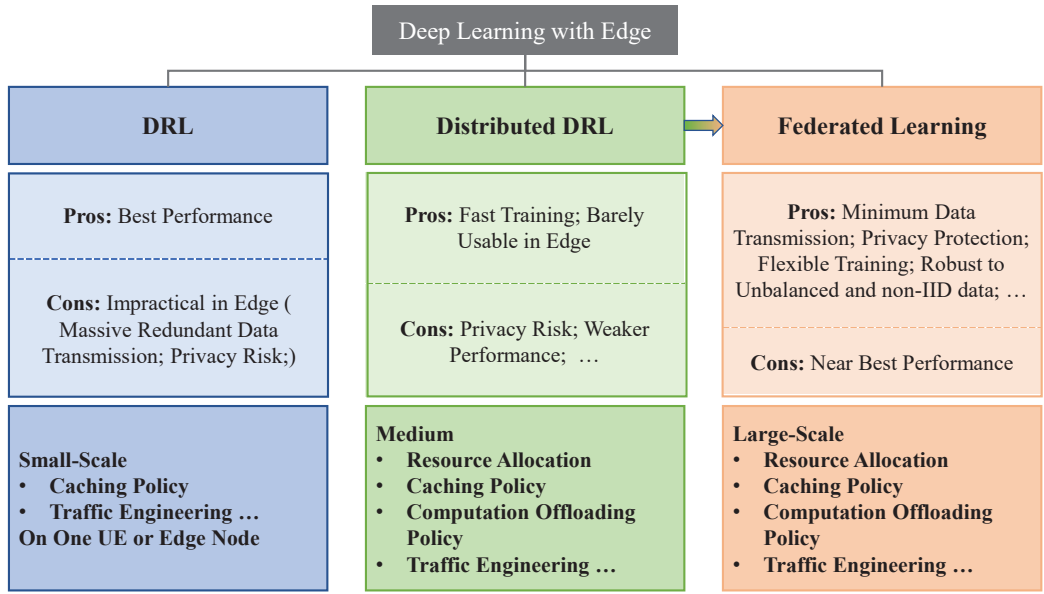}
\caption{Taxonomy of applying Deep Reinforcement Learning in mobile edge system}
\label{fig_3_inedgeaitaxonomy}
\end{figure*}

DRL techniques require intensive computation capacity
for finding the optimal solutions. Particularly, if there are huge
amount of data factors, parameters, and criteria for optimizing the resource
over large-scale MEC systems (an operator's network in cities),
advanced distributed Deep Learning (DL) approaches should be utilized. Hence, thinking about training DRL agents in a distributed fashion efficiently is natural. As the Fig. \ref{fig_3_inedgeaitaxonomy} illustrates, though keeping and training a DRL agent in every UE and edge node can achieve the best performance, it is only practical to apply distributed DRL in the MEC system, because there shall not have enough time and data for training. However, most distributed DRL architectures, described by Fig. \ref{fig_4_tree}(b), could not handle unbalanced and non-IID data and cope with the privacy issues \cite{GOOGLE_FL}. In addition, they usually reduce the performance of DRL agents when UEs and network states are heterogeneous. Therefore, Federated Learning (FL) is introduced in this paper for training DRL agents in the MEC system on account of the fact it could deal with several key challenges below, which differentiates it from other distributed DL approaches \cite{GOOGLE_FL}:
\begin{itemize}
\item \textbf{Non-Independent and Identically Distributed (Non-IDD):} The training data (transition memories in DRL) on the UE is based on the wireless environment it experienced and its own computation capability and energy consumption. Hence, any individual training data of a UE will not be able to represent the training data of all UEs. In FL, this challenge could be met by merging the updates of models with FedAvg in \cite{GOOGLE_FL}.
\item \textbf{Limited communication:} UEs are often unpredictably off-line or allocated with poor communication resources. Nevertheless, using additional computation could decrease the consumption of communication rounds needed to train a model. In addition, FL only asks a part of clients, in one round, to upload their updates, which handles the circumstance where clients are often unpredictably off-line.
\item \textbf{Unbalanced:} Some UEs may have more computation tasks to be handled and some may experience more states of mobile networks, resulting in varying amounts of training data among UEs. Also, this challenge could be coped with the FedAvg algorithm.
\item \textbf{Privacy and Security:} The information needs to be uploaded for FL is the minimal update necessary to improve the DRL agent. Further, techniques of secure aggregation and differential privacy could be applied naturally, which could avoid that privacy-sensitive data are contained in local updates. Nonetheless, the privacy and security is not our focus in this work, more information about these issues could be found in references of \cite{GOOGLE_FL}
\end{itemize}

\subsection{Integration of Federated Learning within Edge for In-Edge AI}

Between AI in cloud and AI in UEs, there is an ``edge''. With the aid of FL framework, edge nodes equipped with AI computation in ``edge'' could combine the cloud and massive UEs together and form a powerful AI entity with strong cognitive ability provided by massive UEs and edge nodes. Throughout this envisioned architecture, each edge node can support AI tasks on system level dynamically, not only for its own but for the global optimization and balancing of the whole MEC system.

Use cases of edge caching in Section \ref{sec:AIForEdgeCaching} and computation offloading in Section \ref{sec:AIForEdgeComputing} are taken to represent In-Edge AI of integrating FL.

\subsubsection{From Edge to Cloud}
Considering edge caching in the MEC system, the DRL agent employed in an edge node makes decisions on caching appropriate content according to requested contents of UEs dynamically. However, space-time popularity dynamics of requesting contents put forward the requirement for collaborated edge caching. Thus, the cloud server belongs to MNOs could be the central server for coordinating the edge nodes. Among all edge nodes, each of them keeps a DRL agent and updates it by its own local training data.

\subsubsection{From UEs to Edge}
In the scenario of computation offloading, each UE shall decide whether one computation task should be offloaded to edge nodes, offloading the task via which wireless channel and the energy consumption according to the inference result of the DRL agent in it. Though UEs such as mobile phones, industrial IoT devices, and smart vehicles are able to perform some AI computation, the computation capability and the energy consumption still limit their abilities in AI computing (DRL training on large-scale data). Therefore, we proposed to use all edge nodes in ``edge'' as an integral server for coordinating massive UEs covered by them. By virtue of this scheme, UEs with relative weak computation capability could be able to hold a complex DRL agent.

\begin{figure*}[!!!!!!!!!!!!!!hhhhhhhhhht]
\centering
\subfigure[Centralized DRL]
{\includegraphics[width=2.4 in]{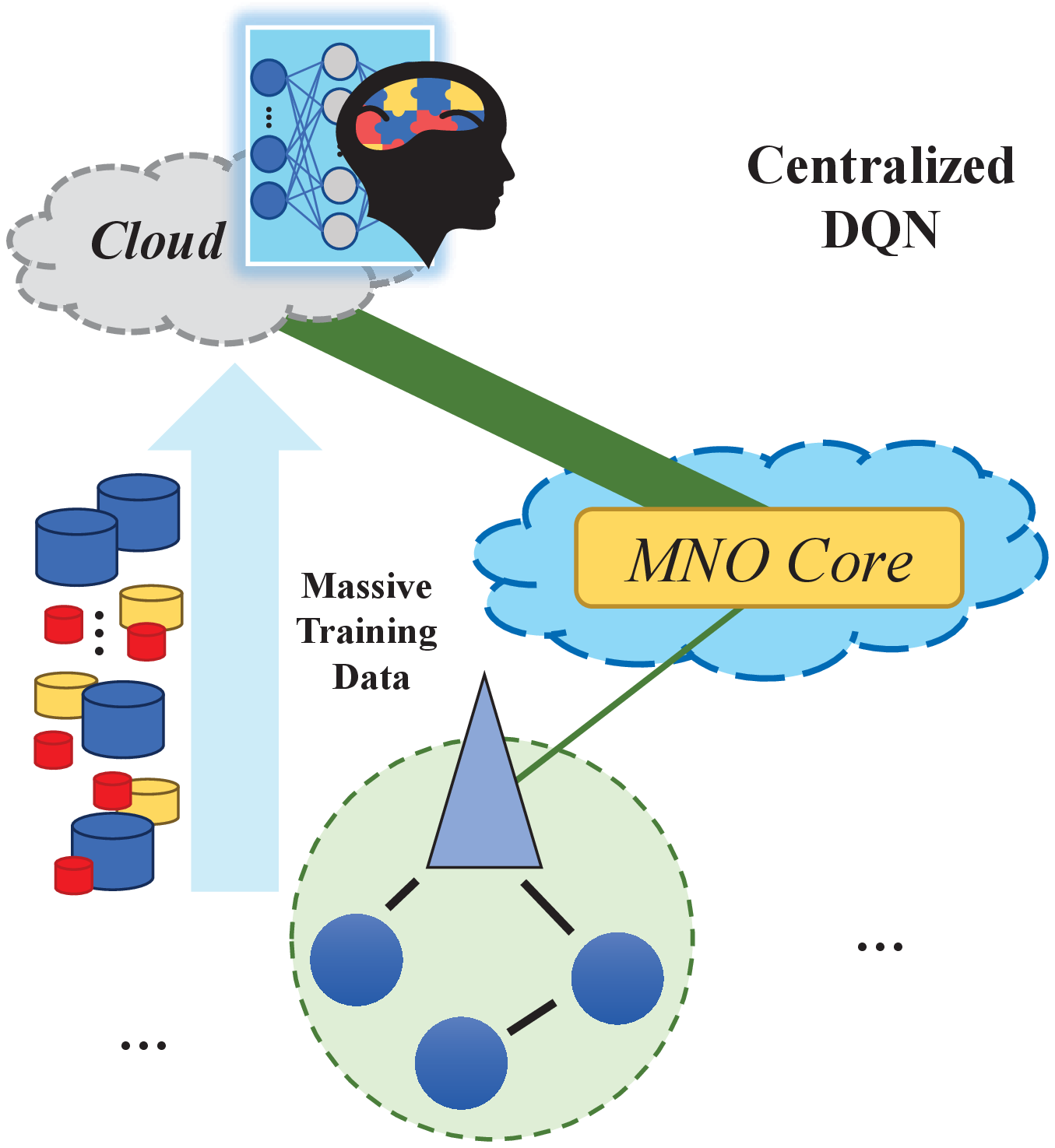}
\label{fig_4_dqnddqnfl_1.eps}
}
\subfigure[Distributed DRL]
{\includegraphics[width=3.1 in]{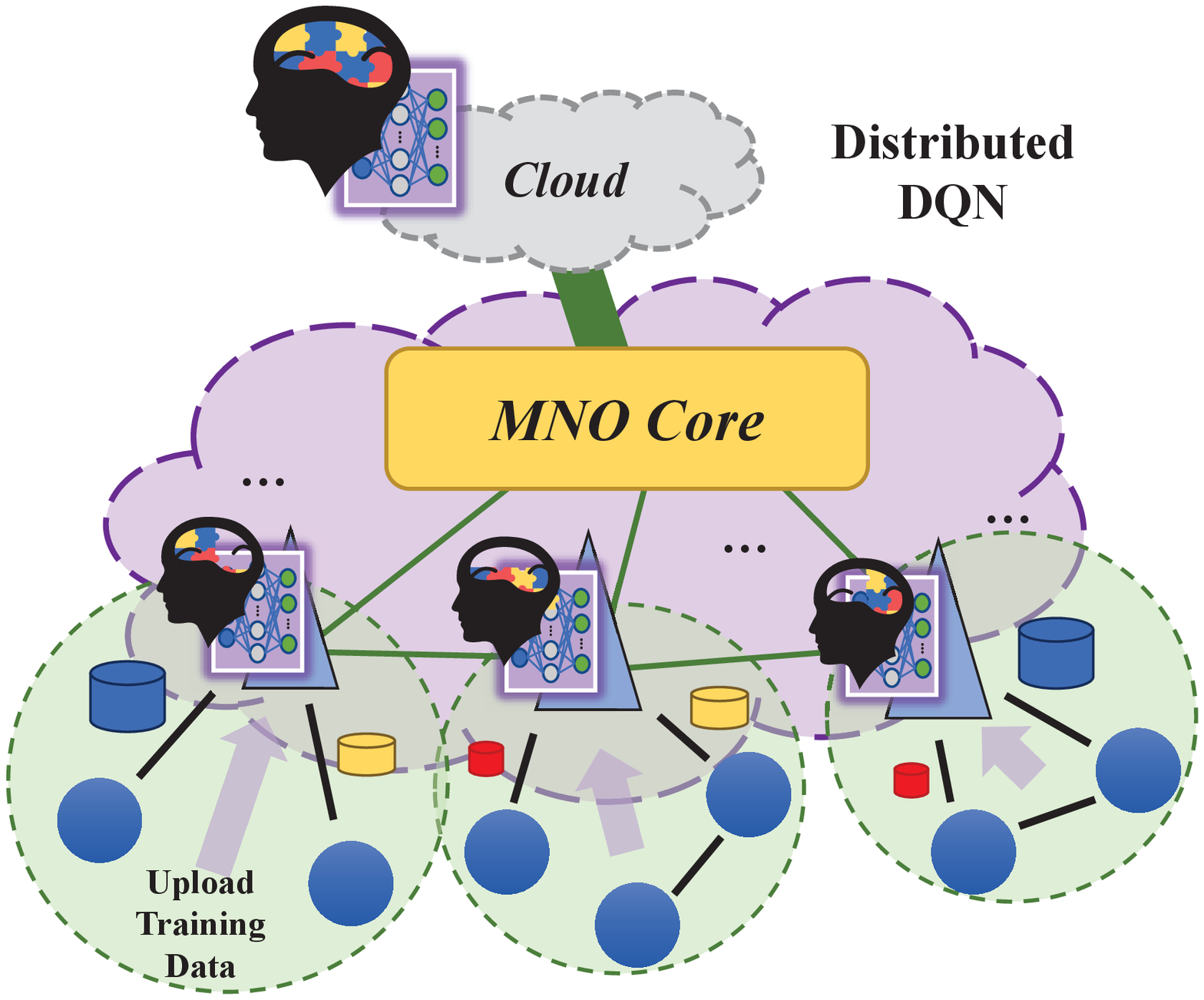}
\label{fig_4_dqnddqnfl_2.eps}
}
\subfigure[DRL with Federated Learning]
{\includegraphics[width=7.0 in]{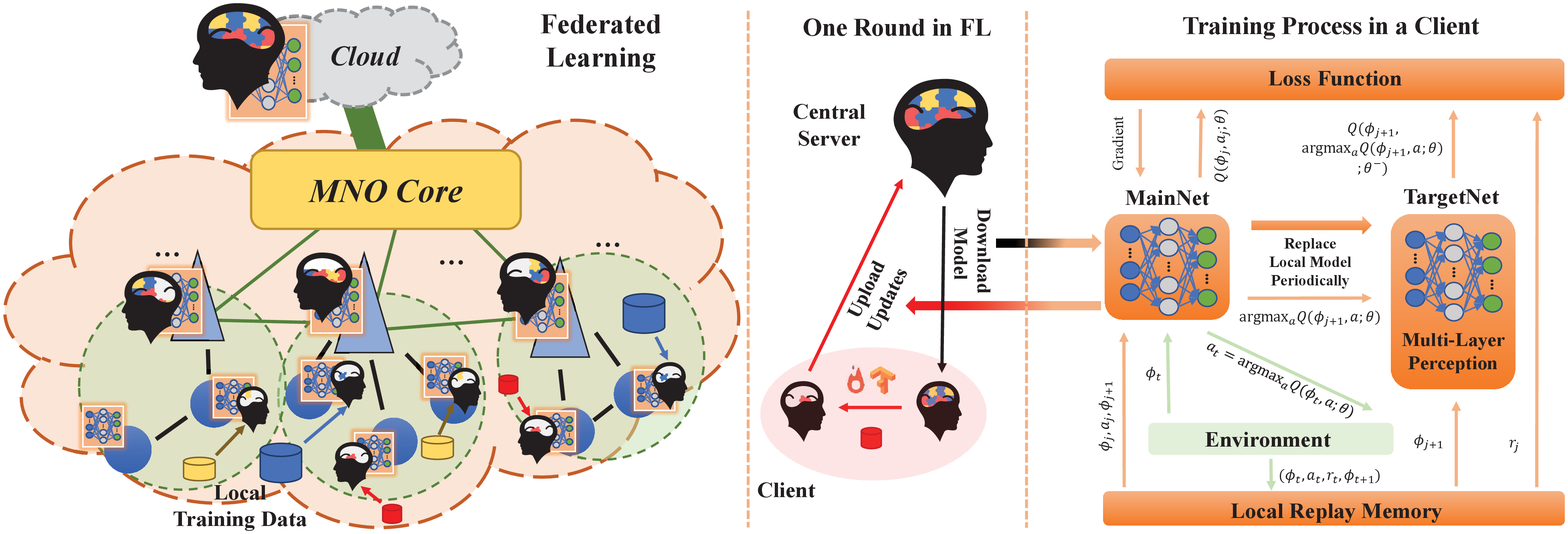}
\label{fig_4_dqnddqnfl_3.eps}
}
\caption{Centralized DRL v.s. Distributed DRL v.s. DRL with Federated Learning over mobile edge system}
\label{fig_4_tree}
\end{figure*}

The methodology of tackling two aforementioned scenarios is similar. For training a general DL model, as depicted in Fig. \ref{fig_4_tree} (c), FL iteratively solicits a random set of clients (distributed devices which train DL model) to 1) download parameters of the DL model from a certain server, 2) perform the training process on the downloaded model with their own data, and 3) upload only the new model parameters to the server, which aggregates uploaded updates of the client to further improve the model. The process inside an individual client is also illustrated in Fig. \ref{fig_4_tree}(c), where symbols are taken as the same version in \cite{BOOK_RL_INTRODUCTION} for accessible reading. Specifically, the client trains its model, which is download from the central server before, based on local training data and upload the updated weight of MainNet backward when it is available.

To summarize, FL enables resource-constrained edge to compute devices (include UEs and edge nodes) to learn a shared model, while keeping the training data local.

\subsection{Pros and Cons of Federated Learning for In-Edge AI}
The core benefit of FL lies in distributing the quality of knowledge across a large number of devices without necessarily centralizing the training data. Extended by this core benefit, several particular benefits are brought further in the MEC system.

\begin{itemize}
\item \textbf{Cognitive:} Server-side proxy data is less relevant than on-device data. In the MEC system, massive UEs and edge nodes could act as perceptrons and acquire various, abundant and personalized data for updating the global DL model. On the perspective of UEs, these data could include the quality of the wireless channel, the remanent battery life and the energy consumption, the immediate computation capability and so on. Concerning edge nodes, the cognitive data could include the computation load, the storage occupation, the number of wireless communication links, the task queue states waiting for handling, and etc. Thus, using FL based on these raw data instead of centralized DL renders the MEC system more cognitive.
\item \textbf{Robust:} FL inherently could address key issues about the availability of UEs and edge nodes and the unbalanced and non-IID data. Consequently, the performance of In-Edge AI shall not be easily affected by the unbalanced data and sometimes the poor communication environment. Further, its ability of handling non-IDD data allows massive UEs in different wireless environments to train their own DL model without concerning about the negative effect.
\item \textbf{Flexible:} In FL, additional computation could be used to decrease the number of communication rounds to train a model. An effective way to add computation is increasing computation per UE, i.e., adding more local SGD updates per round. Hence, a trade-off between computation and communication existed. Specifically, powerful (both in computation and energy) UEs could decide to perform more mini-batches in training for further decreasing the communication cost and vice versa.
\end{itemize}

Certainly, due to the limitation of federated optimization, FL could not outperform the centralized DL trained model, but could still achieve the near best performance, which will demonstrated in Section \ref{sec:evaluation}.

\subsection{Practicability Discussion}
\label{sec:practicability}
To some extent, our proposed In-Edge AI is a future-oriented concept. We envision a near future where most of the mobile devices, particularly smartphones, are endowed with the ability of not only inferring but also training the Deep Learning model. As is well known, even the state-of-the-art edge chips, such as Edge TPU (brought by Google and powered by TensorFlow Lite), could only support elementary training processes in DL. Therefore, the practicability should be discussed with considering both practical deployment and delay requirement.

\subsubsection{Deploy Challenges}
Learning takes long time of training as well as inferring according to the required accuracy level. Obviously, the DRL model should not be deployed directly while setting weights of neural networks at random. Otherwise, the MEC system will be paralyzed because the DRL model could only make random decisions at preliminary exploration.

Nonetheless, this may be solved if the DL model is not trained from scratch. We are now working on using transfer learning to boost the training process in MEC systems. The basic idea is to simulate the wireless environment and requests of UEs. Just as evaluating and adjusting the antenna settings in a simulated testbed, the simulated environment is used to train an off-line Deep Reinforcement Learning agent. Then, the established DRL model could be distributed to mobile UEs.

\subsubsection{Delay Requirement}
Learning always takes long time of training as well as inferring according to the required accuracy level. However, if using transfer learning, there shall be less time or computation consumption on the side of training.

Based on the pre-established DRL, the requirement of training (not inferring) is alleviated, i.e., the DRL agent could reach the satisfied accuracy level after a small number of mini-batches or even directly inferring. And when the wireless environment or the pattern of UEs' requests is changed, the UE could also take advantage of the established DRL to adjust neural networks in its DRL model.

Unlike using an enormous number of hidden layers and neurons in CNN or RNN, the neural network in DRL is merely a Multi-Layer Perceptron (MLP, shown in Fig. 4(c)). And in our simulated scenarios, the MLP has only one hidden layer with 200 neurons. Therefore, once the model has been trained, the inferring process could be performed quickly due to the low complexity of the MainNet in DRL agents.

Besides, our work \cite{xuchen_aiedge} aims to meet the challenge of running DL on resource-constrained mobile devices. By incorporating this work, the requirement on computation capacity of UEs could be relaxed along with decreasing the inferring delay.

\section{Data-Driven Evaluation of Proof-of-Concept In-Edge AI Framework}
\label{sec:evaluation}

\subsection{Experiment Settings}

For investigating the performance of In-Edge AI with FL, two simulations on edge caching and computation offloading as in Section \ref{sec:AIForEdgeCaching} and Section \ref{sec:AIForEdgeComputing} are presented, respectively. Among all simulations, the time horizon is discretized into time epochs.

We capture Xender's trace for one month (from 01/09/2016 to 30/09/2016), including 9,514 active mobile users, conveying 188,447 content files, and 2,107,100 content requests \cite{JSAC_2018_XIUHUA}, of which the content popularity distribution can be well fitted by a Zipf distribution with $\partial = 1.58$. In edge caching simulations, we use this mined Zipf distribution as the content popularity distribution to generate the content request of UEs, and consider the cooperation of $6$ edge nodes. Once an edge node receives a request from UE, the local DRL agent in it will make a decision to cache which content or not cache, and then obtain the reward of this action for its own training. In edge caching, the cloud server belongs to MNOs is the central server for coordinating the edge nodes.

On investigating the capabilities of DRL coupled with FL over the MEC system for computation offloading, we suppose the whole bandwidth $\omega = 5$ MHz of an edge node is divided into $10$ wireless channels, and take $10$ UEs as the clients in FL framework to individually train their DRL agents and merge them among edge nodes. The channel gain states between the UE and the edge node are from a common finite set, which quantifies the quality of the wireless channel into $6$ levels. Throughout the simulation, the number of tasks generated on each UE follows Bernoulli distribution with average arrival rate $\lambda_{\mathrm{T}}$ per time epoch.

As for the DRL settings in UEs, edge nodes, and cloud, we choose the Double DQN algorithm and select tanh as the activation function and Adam optimizer. We used a single-layer fully-connected feedforward neural network, which includes $200$ neurons, to serve as the target (TargetNet) and the eval (MainNet) Q network. Other parameter values in Double DQN are set as follows: replay memory capacity $M=5000$, mini-batch size $B=200$, discount factor $\gamma=0.9$, exploration probability $\epsilon=0.001$, learning rate $\zeta=0.005$ and the period of replacing the target Q network is $\phi=250$. In addition, to build a baseline for the performance of DRL agent with FL, we construct a centralized DRL agent for comparison and it is assumed to be able to receive all data used for reinforcement learning.

\subsection{Evaluation Results}

To elucidate the performance of our In-Edge AI framework, experiments on edge caching and computation offloading are carried out under various settings.

\begin{figure*}[t]
\centering
\includegraphics[width=17cm]{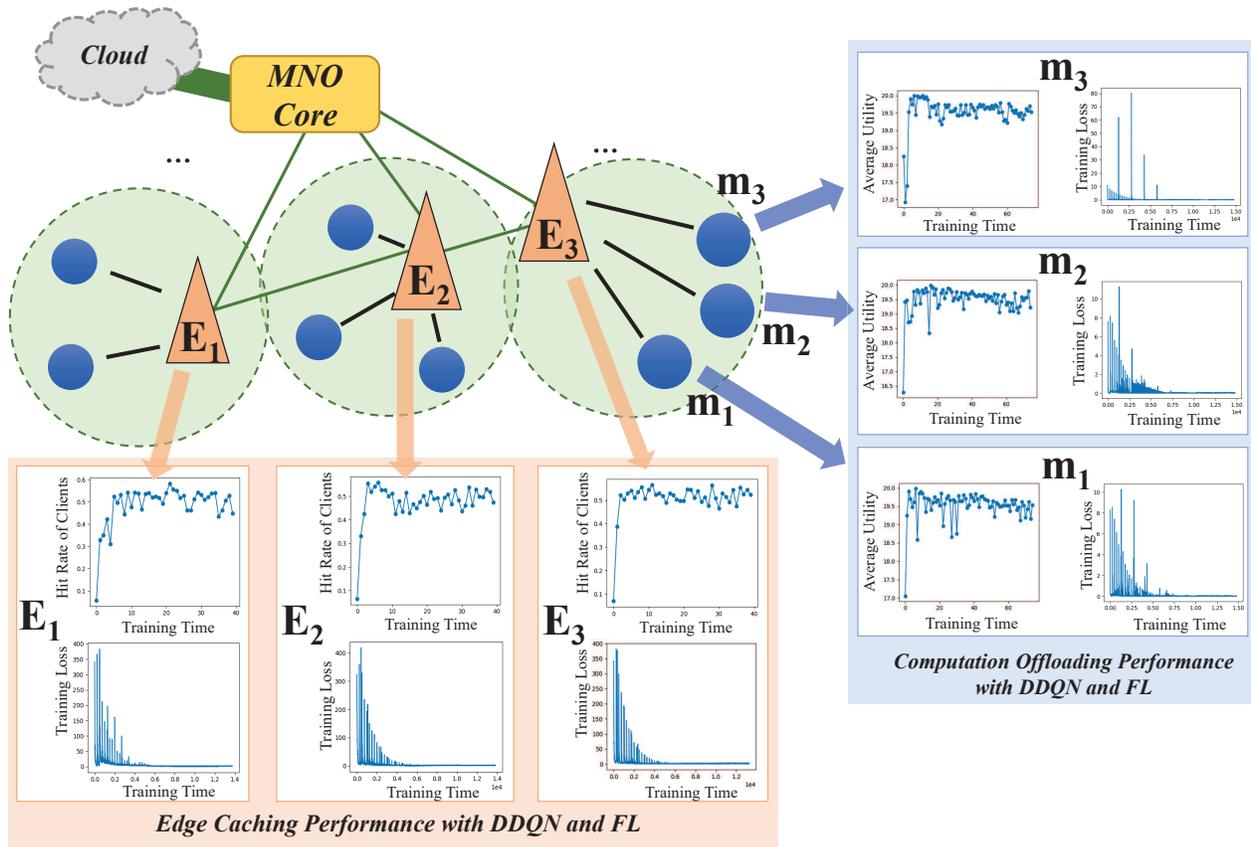}
\caption{Capability of Federated Learning in edge caching and computation offloading}
\label{fig_5_results}
\end{figure*}

Fig. \ref{fig_5_results} depicts the performance of DDQN with FL both on edge caching and computation offloading. Three edge nodes ($\mathrm{E}_{1}$, $\mathrm{E}_{2}$ and $\mathrm{E}_{3}$) and three UEs ($\mathrm{m}_{1}$, $\mathrm{m}_{2}$ and $\mathrm{m}_{3}$) are chosen for exhibiting the capabilities of their own DRL agents. On the perspective of edge caching, $\mathrm{E}_{1}$, $\mathrm{E}_{2}$ and $\mathrm{E}_{3}$ are chosen to cache contents, and hit rates of them are improving and finally maintain within a certain range along with the process of training DRL agents.
In the simulation of computation offloading, the average utilities of three UEs $\mathrm{m}_{1}$, $\mathrm{m}_{2}$ and $\mathrm{m}_{3}$ are increasing with the decreasing of training loss, and also maintain within a certain range at last. This means that the efficiency of handling offloaded tasks in the MEC system is improved.

\begin{figure}[!!!!!!!!!!!!!!hhhhhhhhhht]
\centering
\subfigure[Performance comparison in edge caching]
{\includegraphics[width=4.8 cm]{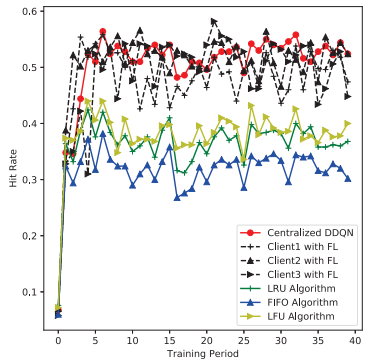}
\label{fig_6_results_1}
}
\subfigure[Performance comparison in computation offloading]
{\includegraphics[width=4.8 cm]{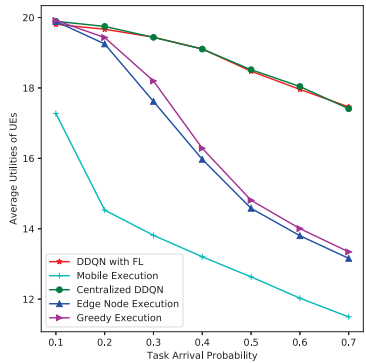}
\label{fig_6_results_2}
}
\subfigure[Detailed performance comparison in computation offloading]
{\includegraphics[width=4.8 cm]{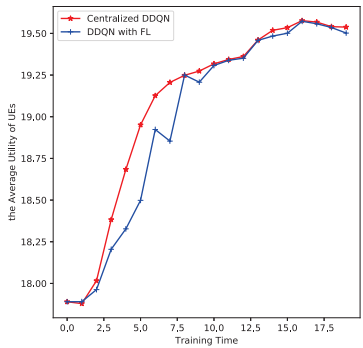}
\label{fig_6_results_3}
}
\subfigure[Comparison of transmission cost]
{\includegraphics[width=4.8 cm]{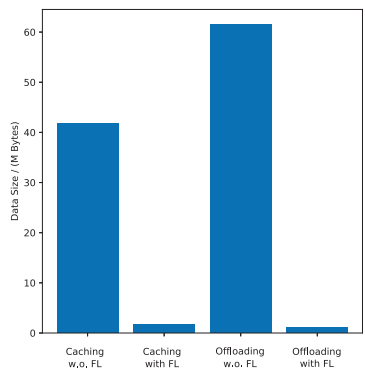}
\label{fig_6_results_4}
}
\caption{Performance Evaluation of Double DDQN with and without Federated Learning}
\label{fig_6_results}
\end{figure}

Further, Fig. \ref{fig_6_results} gives the details of performance comparison as follows.
\begin{itemize}
  \item[1)] In Fig. \ref{fig_6_results_1}, the edge caching performance of the DDQN with FL is compared to the centralized DDQN, and some other caching policies, such as known LRU (Least Recently Used), LFU (Least Frequently Used) and FIFO (First in, First out). It can be easily seen that the specific performance of FL is near close to the results of centralized DDQN in terms of achieved hit rates of $3$ clients (edge nodes), and outperform LRU, LFU, and FIFO.
  \item[2)] In the experiment on computation offloading, the performance of the DDQN with FL is compared to the centralized DDQN, and another three baseline computation offloading policies, i.e., mobile execution, edge node execution, and greedy execution. Here, mobile execution, edge node execution and greedy execution mean that the UE processes all computation tasks locally, all computation tasks are offloaded from the UE to edge nodes, and the UE makes decision on executing a computation task locally or offloading it to edge nodes with the aim of maximizing the immediate utility, respectively. In Fig. \ref{fig_6_results_2}, it can be observed that the DDQN with FL achieves the near performance of centralized DDQN and is superior to another three baseline policies;
  \item[3)] On investigating the detailed performance in training process, we assume that the capability of wireless communication is not the hinder, i.e., both massive training data for centralized DDQN and lightweight model updates for DDQN with FL can be uploaded to the target position within a discretized time epoch. As illustrated in Fig. \ref{fig_6_results_3}, the performance of centralized DDQN is better than that of DDQN with FL at the beginning of training. However, once the model merging of FL has been processed several times, the performance of DDQN with FL becomes near to that of centralized DDQN. Certainly, if the client wishes to obtain the expected performance using DDQN with FL, it must take time to wait for the model merging, i.e., exploiting training results of other clients. Nonetheless, this experiment assumes an ideal wireless environment. In practice, massive training data are actually unable to be uploaded without any delay. Therefore, performing DDQN with FL is more practical in MEC systems, at least for now when the wireless resource is also the major consideration.
  \item[4)] We also gather statistics of the total wireless transmission until the DRL agent is well trained both in the scenario of edge caching and computation offloading. In the framework of FL, every client only needs to upload the update of its model. Without FL framework, viz., using centralized DRL, UEs must upload the whole training data via wireless channels and thus consume more communication resources, which is demonstrated by  Fig. \ref{fig_6_results_4}.
\end{itemize}

With no doubt, FL must trade something for its advantages. Specifically, due to the fact that the coordinating server in FL only executes tasks of merging updates instead of taking over the whole training, the computation load of clients is inevitably heavier on account of the local training process. This will cause more energy cost on UEs or burden the computation of both UEs and edge nodes, and these issues are still open questions.

\section{Opportunities and Challenges}
\label{sec:opportunitiesandchallenges}
We hereby discuss a few essential promising research directions of In-Edge AI,
and highlight several pending problems from the perspective of elaborating and widening the usage of AI and Edge Computing.

\subsection{Accelerating AI Tasks by Edge Communication and Computing Systems}
Aforementioned sections is to optimize the edge of mobile communication systems by AI techniques,
but it is also important to exploit specialized methods of optimizing learning computation tasks
by the support of edge \cite{xuchen_aiedge}.
DRL can be treated as a type of specialized edge computing task at the service-level.
How to find correct edge nodes to collaborate, and how to allocate appropriate resources,
for the large amount of AI tasks with various priorities, deadlines, scales, different requirement
on CPU, memory and so on are also indispensable.
Interestingly, this is the reverse direction of the aforementioned Federated Learning,
but also a kind of application of the ``In-Edge AI''.
Potential research topics may include
game theoretic algorithms for the competition of edge nodes and mobile users
over the multi-dimensional resource for AI acceleration by the edge,
and the dynamic and adaptive splitting of AI tasks becomes quite challenging.

\subsection{Efficiency of In-edge AI for Real-time Mobile Communication System towards 5G}

In 5G, the defined Ultra-Reliable Low-Latency Communications (URLLC) scenarios strongly
desire very small delays and high reliability of the mobile system.
However, general deep learning-based optimization and prediction schemes take quite long running time
of recursions for converging to the results, which is inappropriate for mobile edge systems;
particularly the system-level edge computing tasks desiring rapid responses in the scale of millisecond.
In order to manage the guaranteed QoS of delays and bandwidth for caching, communication and computation,
In-Edge AI should be able to provide differentiated support for various types of services,
and fine-grained collaborative scheduling of the AI tasks (or split ones) over the edge nodes and mobile devices
should be accelerated in nearly real-time, which is critical but remain unsolved in current literature.
Furthermore, the integration of Federated Learning framework with AI chipset hardware should be explored as well.

\subsection{Incentive and Business Model of In-Edge AI}

In-Edge AI is a tight confederation among mobile operators, SPs/CPs, and mobile users,
while capable entities may contribute more to the overall optimization of the edge caching and computing tasks,
but a large number of UEs may rely on the AI of edge nodes and other UEs.
Also, AI computation requirements from SPs and CPs should be satisfied across edge nodes of different mobile operators.
And hence, more reliable and accurate design of the incentive framework of In-Edge AI become interesting,
which should further provide a loop of the digital copyrights of the content and services.
Blockchain techniques may be integrated into In-Edge AI framework \cite{blockchain_edge},
but how to distribute the huge computation load of the proof of work over the edge system
and how to evaluate the contribution of In-Edge AI computation on heterogenous scenarios
are still unexplored.

\section{Conclusions}
\label{sec:sonclusion}
In this article, we discussed the potential of integrating the Deep Reinforcement Learning techniques and Federated Learning framework with the mobile edge system and optimizing the mobile edge computing, caching and communication with it.
We perform experiments on investigating the scenarios of edge caching and computation offloading in mobile edge system, and
the ``In-Edge AI'' is evaluated and proved to have the abilities to achieve near-optimal performance. For our future work, we concentrate on not only optimizing learning computation tasks by the support of edge
but also scheduling the AI tasks over the edge nodes and mobile devices in a fine-grained and collaborative manner.

\section*{Acknowledgement}



\begin{biography}
Xiaofei Wang (xiaofeiwang@tju.edu.cn)
is currently a professor in Tianjin University, China.
He received M.S. and Ph.D degrees from the
School of Computer Science and Engineering, Seoul National
University in 2008 and 2013 respectively. He received the
B.S. degree in the Department of Computer Science and Technology, Huazhong University
of Science and Technology in 2005. He is the winner of the
IEEE ComSoc Fred W. Ellersick Prize in 2017.
His current research interests are social-
aware multimedia service in cloud computing, cooperative
backhaul caching and traffic offloading in mobile content-
centric networks.
\end{biography}

\begin{biography}
Yiwen Han [S'18] (hanyiwen@tju.edu.cn)
is a Ph.D student in Tianjin University, China.
He received the B.S. and M.S. degrees from Nanchang University and Tianjin University, China, in 2015 and 2017, respectively.
His current research interests include edge computing, wireless communication, and machine learning.
\end{biography}

\begin{biography}
Chenyang Wang (chenyangwang@tju.edu.cn)
received his B.S. and M.S. degrees from Henan Normal University, Xinxiang, Henan province, China. He is now a Ph.D student in the School of Computer Science and Technology at Tianjin University. His research interests include Mobile Edge Computing, Caching, Offloading and D2D wireless networks.
\end{biography}

\begin{biography}
Qiyang Zhao
is currently a senior research engineer at Huawei Technologies, China.
He received B.S. and Ph.D degrees from Xidian University and University of York in 2005 and 2013 respectively.
He is specialized in end-to-end network slicing, radio resource management, control and user planes, network orchestration, optimization, reinforcement learning, transfer learning, neural network and data analytics, plus comprehensive skills in mathematical modeling, system level simulation and large-scale prototype development.
\end{biography}

\begin{biography}
Xu Chen [M'12] (chenxu35@mail.sysu.edu.cn)
received his Ph.D. degree in information engineering
from the Chinese University of Hong Kong in 2012. He was a
postdoctoral research associate with Arizona State University,
Tempe, from 2012 to 2014, and a Humboldt Fellow with the
University of Goettingen, Germany, from 2014 to 2016. He is
currently a professor with the School of Data and Computer
Science, Sun Yat-sen University, Guangzhou, China. He received
the 2017 IEEE ICC Best Paper Award, the 2014 IEEE INFOCOM
Best Paper Runner-Up Award, and the 2014 Hong Kong Young
Scientist Runner-Up Award.
\end{biography}

\end{document}